*Article*

# Fisher-Rao Distance Detects Shifts in Kinematic Profiles under Cognitive Load

**Joseph Vero [1,*] and Elizabeth Torres [1,2,3,*]**

[1] Rutgers the State University of New Jersey, USA, Psychology Department; joseph.vero@rutgers.edu
[2] Rutgers University Center for Cognitive Science; ebtorres@psych.rutgers.edu
[3] Rutgers University Center for Biomedicine Imaging and Modelling, Computer Science Department

* Correspondence: ebtorres@psych.rutgers.edu, joseph.vero@rutgers.edu

**Abstract**

Motor control research involves the study of movement kinematics derived from the positional trajectories that complex motions describe. In natural, unconstrained motions requiring cognitive and memory processes in real time, the temporal speed profiles are not bell-shaped, may have multiple maxima and the peaks distribution is best fit by the continuous gamma family with two parameters, the shape and the scale. As the stochastic processes described by complex motion trajectories are non-stationary, the gamma shape and scale parameters describing them span stochastic trajectories. These points live on a curved surface where Euclidean distance depends on the arbitrary choice of parameterization. We adopt the Fisher-Rao distance (the geodesic length on the gamma manifold) as a coordinate-free metric for movement fluctuation signatures and present a robust numerical solver that converges across the full range of empirically observed parameters. We demonstrate the metric on a tablet-based digitized Trail Making Test in healthy adults, comparing micromovement fluctuations under low and high cognitive load. Load displaced every participant's fluctuation signature but with no shared direction. Instead, participants converged toward a common operating regime, with those farthest from it at baseline moving the most. Cognitive load thus contracts individuality in motor fluctuations rather than shifting the population uniformly.

## 1. Introduction

Scientists typically quantify movements embedded in behavior for two reasons: to measure the differences between two different behaviors, or to measure the variation between two realizations of the same behavior. The first use-case has a straightforward application. Differentiation of movements has useful applications in clinics, robotics, and in automated action segmentation in sports and kinesiology applications. Differentiation within movements, however, requires a more nuanced rationale. When quantifying variations of different realizations of otherwise equivalent movements, there are a few sources that contribute to measured variability. Classically, the canonical delineation for variability subtypes is “task-relevant” vs. “task-incidental” [1-6]. More recently, researchers have found that axes of variability in the null space of the movement (traditionally considered the "task-incidental") are in fact core to learning the movement, both in humans [5,7] and non-humans [6,8,9]. However, it is important to differentiate this useful variability from variability that is symptomatic of degradation of performance, either in planning or in execution.

Variability contributes to the notion of noise and noise is ever present in the nervous system [10]. One source of noise that has been studied extensively is cognitive load via a dual task. Most relevantly, dual tasks have shown differences in drawing velocity between people with mild cognitive impairment (MCI) and healthy controls [11]. This approach can differentiate between conditions, like other paper-based tasks for screenings for dementia, but, importantly, these tasks are not fine-grained enough to resolve minute differences between subtypes of dementia [12]. Additionally, these tasks cannot give a proper metric for severity, nor can they characterize a "trajectory" to quantify efficacy of treatment apart from coarse metrics like mean velocity.

Mean velocity, as a motivating example, is an improper target to quantify efficacy of interventions, as it may decrease when neuromotor function is increasing, or worse, it may increase due to the patient developing compensatory mechanisms [13]. Furthermore, across participants, the ranges of mean speed obtained from mean magnitude of velocity are impacted by anatomical disparities, a fact that is rarely considered in cohorts of people with different limb and body lengths. These factors make simple metrics like mean velocity antithetical to a proper metric of measuring the restoration of neuromotor function. Making a metric that is useful, in the sense of being portable, properly standardized, and general, requires a theoretically meaningful notion of distance for what is being measured, and accurate measurement requires a ruler matched to the geometry of the problem.

To build a measurement of movement profiles, we use the methods developed in [14], characterizing fluctuations of velocities on the plane of the continuous family of gamma distribution parameters. Gamma parameters, however, live on a curved surface, so any distance between pairs of $(\alpha, \beta)$ points must take that curvature into consideration (where $\alpha$ represents the shape and $\beta$ represents the scale or dispersion of the distribution). A fixed step in $(\alpha, \beta)$ does not change a distribution in a constant way. Two distributions that differ by $(d\alpha, d\beta)$ could have wildly different distinguishability, based on the initial position of $(\alpha, \beta)$. A valid ruler would take this distinguishability into consideration when defining a notion of distance between distributions, independent of the chosen parameterization. This distance allows us to quantify the physical kinematic effects of cognitive load via set shifting and is placed on a common scale which allows comparison within and between people. This gives us a theoretically appropriate tool that provides distances between movement types and within movement types, capturing small yet important nuances across different contexts and situations.

Here, we apply this framework to a digitized, tablet-based variant of the Trail Making Test, in which participants trace between sequentially numbered targets (TMT-A), then between alternating alphanumeric targets (TMT-B), imposing a set-shifting cognitive load. Using the Fisher-Rao distance (FR distance) [15], we measure how far the cognitive load displaces each person's signature of kinematic fluctuations on a common scale. We find that TMT-A shows individually distinct kinematic signatures between people, which converge toward a common regime under TMT-B. This suggests that the required set-shifting results in a reduction of individuality in the context of kinematic fluctuations in healthy participants.

## 2. Materials and Methods

Drawing data was collected from 19 healthy adults (mean age 28.8 ± 8.6, 9 females and 10 males) with no reported history of mobility impairment, cognitive impairment, or neurological injury, providing a reference baseline against which clinical populations can be compared in future work. Recruitment and consent procedures were approved by IRB protocol # Pro2020000154. Details are described in full in [16] and briefly summarized below.

Each participant performed a connect-the-dots tracing task adapted from the Trail Making component of the Montreal Cognitive Assessment (MoCA), itself derived from the Reitan Trail Making Test [17]. We refer to this digitized trail making test as dTMT. For each trial, 12 numbered targets were displayed at randomly generated, non-overlapping positions (hit radius ≈ 12.5 px), and the participant was instructed to connect the targets in order from start to finish in a single continuous drawing. We placed no other constraints on the shape of the connecting paths, did not prohibit path crossings, and imposed no time limit, deliberately avoiding any instruction that could bias the kinematics of the movement.

Cognitive load was manipulated within subject by varying the target sequence. In the low-load condition the targets formed a purely numeric sequence (1-2-3-4-5…); in the high-load condition they formed an alternating alphanumeric sequence (1-a-2-b-3-c…), requiring participants to maintain and switch between two ordered sets. This numeric-versus-alphanumeric contrast parallels the standard Trail Making Test A/B manipulation of executive demand.

Movements were recorded on an M1 11″ iPad Pro using a second-generation Apple Pencil, with pen-tip position sampled via PencilKit at a measured sampling rate of 240 Hz. Only the time series of (x, y) position and the associated timestamps were retained for analysis; available pen orientation and force channels were omitted to keep the data comparable to prior kinematic work [18]. Full details of the data-acquisition application are given in [16].

For this procedure, each participant performed dTMT-A then dTMT-B, each with 12 targets. Participants were instructed to connect the dots without lifting the Apple Pencil, as lifting the pen resets the trial, to ensure that all the kinematics from the trial are captured.

*2.1 Data Preprocessing*

Each segment sub-trail (the position time-series travelling from target i to target i+1) is extracted. From those extracted positions, we perform backward finite difference to approximate x and y velocity (sampling rate of the pencil is reported and validated as 240 Hz). Afterwards, we take the Euclidean norm of vx and vy to get the velocity of the pen during the trial. To mitigate quantization error of the iPad, we smooth the velocity using a 13-sample unit-sum Gaussian window (gausswin in MATLAB R2025b). This gives us a proper velocity time series to conduct the micromovement analysis. Following [14], we identify the peaks of the speed profile and use maximum likelihood estimation (MLE) to fit the best distribution via MLE, from the continuous gamma family to the normalized peak amplitudes. We fit two parameters, the gamma shape and the gamma scale with 95% confidence intervals. Using these empirically estimated gamma parameters, we obtain the empirical gamma moments (mean, variance, skewness and kurtosis).

We then take the absolute deviation of the speed peaks from the empirically estimated gamma mean and conduct the micromovement analysis on this mean-shifted excursion. From the mean-shifted series, we find all local maxima and minima (using the imregionalmax / imregionalmin functions in MATLAB R2025b, which requires the Image Processing Toolbox). Between each pair of consecutive minima, compute the average speed A and peak P, and compute the normalized micromovement. $MM = \frac{P}{P+A}$. By construction, the normalized MM spikes exist between (0.5-1), as P + A < 2P, in any non-degenerate time series, therefore $\frac{P}{P+A} > \frac{1}{2}$. Similarly, by construction A is non-negative, therefore $P + A \geq P$ and $\frac{P}{P+A} \leq 1$. A schematic of the preprocessing pipeline is given in the Appendix (Appendix A, Figure A1).

*2.2 The Gamma Plane as a Metric Space*

Rather than treating these gamma estimates as a statistic to simply test hypotheses, it is useful to consider them as points in a space, such that we can establish distances, perform clustering, and consider their paths when transitioning between profiles. As a continuous metric, it is feasible and potentially useful to characterize these estimations as a stochastic trajectory to gauge the flexibility of switching between these various profiles. If these gamma parameters represent an underlying cognitive profile, measuring how the parameters change over time may prove useful to understanding strategies underlying various forms of cognitive flexibility above and beyond motor biomechanics. Additionally, it gives us a tool to quantify shifts in kinematics in a principled way, invariant to reparameterization.

Measuring the distance between two probability distributions is already a familiar operation in parametric statistics, even when it is not explicitly named as such. When we standardize a measurement as a z-score, or compare multivariate observations by their Mahalanobis distances, we are implicitly measuring distance not in raw units on a flat $\mu$, $\sigma^2$ parameter plane, but rather in terms of the surface those parameters live on. The distance in that curved parameter space is governed by the Fisher information, and is known as the Fisher-Rao distance [15]. In the case of a z-score, the variance is held constant, so the curvature is constant and zero, collapsing the integral to a closed form solution. In the general Gaussian case, the metric is no longer constant $I(\theta) = \begin{pmatrix} 1/\sigma^2 & 0 \\ 0 & 1/2\sigma^4 \end{pmatrix}$, and curvature is a constant K = -1/2, so the distance is still closed form and straightforward to compute. In the case of gamma distributions, no closed form exists [19], so we must estimate it numerically.

*2.3 Demonstrating Why Fisher-Rao is Necessary*

A gamma distribution can be written in many equivalent coordinate systems: (shape, scale), (mean, variance), and others, and the choice among them is arbitrary. Euclidean distance in parameter space depends on that choice, so it can return contradictory answers about which distributions are alike. Figure 1 makes this concrete with a reference distribution R and two candidates: under the (shape, scale) parameterization, Euclidean distance ranks P as closer to R (1.46 vs. 2.59), but under the (mean, variance) parameterization the ranking reverses, with Q closer (0.34 vs. 1.50). The distributions are unchanged, only the coordinates differ, yet the verdict flips. The Fisher-Rao distance is invariant to reparameterization and returns a single answer (Q closer, 0.75 vs. 3.35) that matches the distributions' actual overlap (Fig. 1, panel A). We therefore adopt the Fisher-Rao distance: the geodesic length on the gamma manifold under this metric, as a coordinate-free ruler, and turn next to computing it.

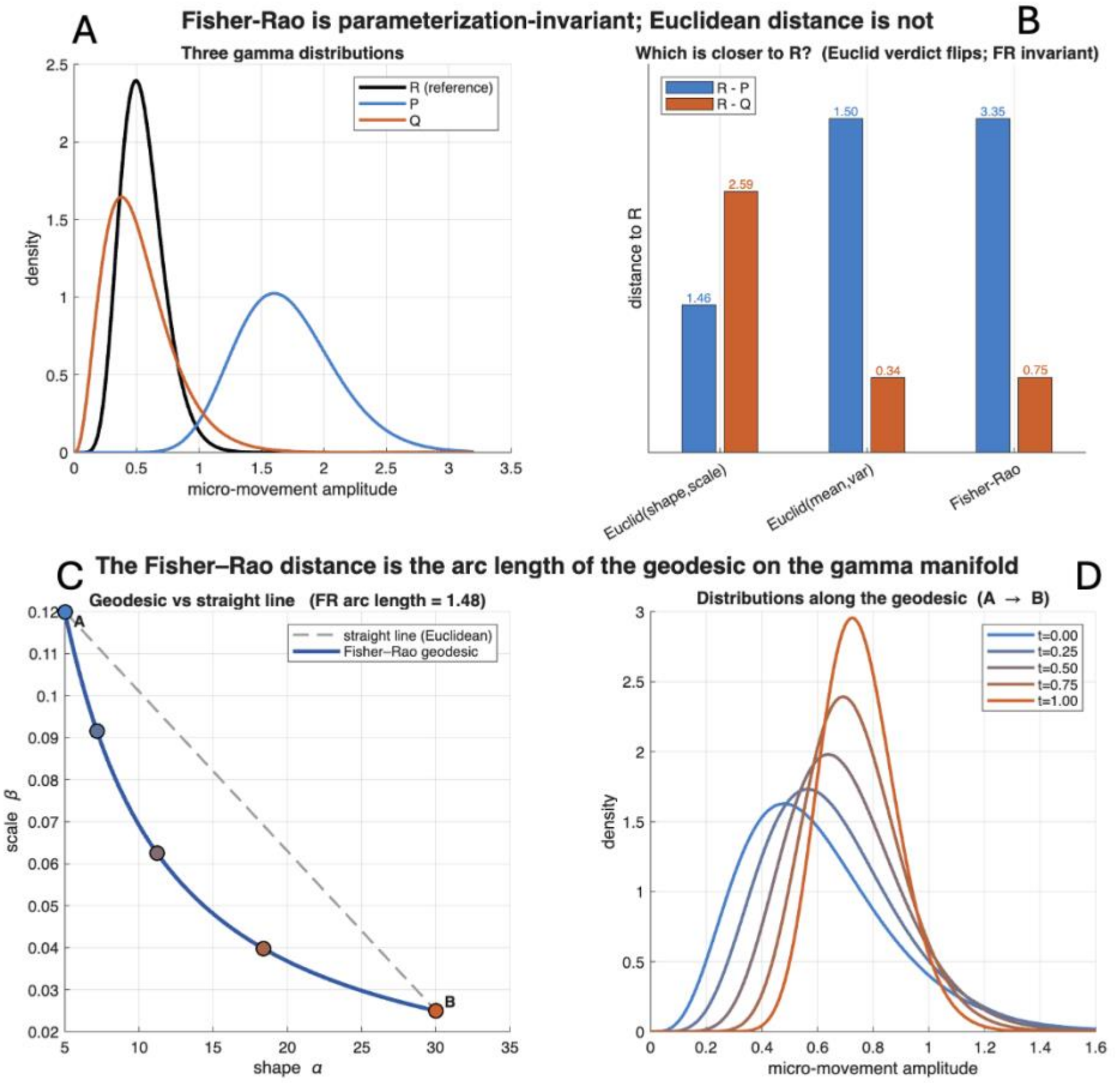

Figure 1 - Consequences of an inconsistent and, in some cases, inaccurate ruler. Panel A and B demonstrate that using arbitrary parameterizations of a distribution can give inconsistent conclusions. Fisher-Rao distance is unique and invariant to reparameterization under sufficient statistics. Chentsov's theorem [20] states that the Fisher information metric is, up to rescaling, the unique Riemannian metric on a statistical manifold. Panel C and D demonstrate that the path travelling along the Euclidean geodesic in flat space differs from the geodesic when travelling along the curved space of gamma parameters. While this bending is less pronounced in the log-log space where we conduct much of the analysis, this only underscores why a consistent ruler is so valuable

### *2.4 Computing Geodesic Arc Length between Gamma Parameter Pairs*

#### 2.4.1 Background and Parameterization

To quantify the dissimilarity between two gamma distributions, we use the Fisher-Rao distance, a metric from information geometry. Under this framework, the two-parameter family of gamma distributions is treated as a curved two-dimensional surface, where each point on the surface corresponds to one distribution and local distances are defined by the Fisher information matrix. The Fisher-Rao distance between two distributions is the length of the *geodesic* connecting them, which is the shortest path across this surface. Unlike commonly used measures such as the Kullback-Leibler divergence, the Fisher-Rao distance is a true metric (symmetric and satisfying the triangle inequality) and is invariant to how the distributions are parameterized.

Closed-form expressions for this distance exist only in special cases (for example, when the two distributions share a common shape parameter), so the general case must be solved numerically. Our implementation follows the approach of [19], who developed the numerical machinery for Fisher-Rao geodesics in the gamma family.

Following that work, we re-express each distribution gamma$(\alpha,\beta)$, with shape $\alpha$ and scale $\beta$, in the coordinates $(\alpha,\eta)$, where $\eta=\log(\alpha\beta)$ is the logarithm of the distribution's mean. In these coordinates the Fisher information metric takes the diagonal form

$$ds^2=\frac{\varphi(\alpha)}{\alpha}\,d\alpha^2+\alpha\,d\eta^2 \qquad \varphi(\alpha)=\alpha\,\psi'(\alpha)-1 \tag{1}$$

where $\psi'$ is the trigamma function. The diagonal (orthogonal) form simplifies the geodesic equations, which are a coupled pair of second-order ordinary differential equations (ODEs) in $\alpha(t)$ and $\eta(t)$ (Eq. 3.4 of Reverter & Oller, 2003).

2.4.2 Solution Strategy

Finding the geodesic between two given distributions is a boundary value problem: the path's two endpoints are fixed, but its initial direction is unknown. The solver approaches this with a three-stage strategy, where each stage acts as a fallback for the previous one.

Stage 1: Shooting with an analytically derived starting guess

The boundary value problem is converted into a sequence of *initial value problems*: we guess the initial velocity (direction and speed) of the geodesic at the first endpoint, integrate the geodesic ODEs forward, and measure how far the resulting endpoint lands from the target. The initial velocity is then refined by Newton's method until the endpoint mismatch falls below a relative tolerance of $10^{-7}$.

Two features make this stage robust. First, the initial guess is not arbitrary: the gamma manifold can be locally approximated by a hyperbolic (Poincaré half-plane) geometry, for which geodesics are known in closed form. We compute the starting velocity from two such approximations (corresponding to embedding curvatures $b=2$ and $b=4$) and average them (Eqs. 3.7-3.10 of [19]); supplementary guesses derived from the geodesic's first integrals are also tried if the primary guess fails. Second, the sensitivity of the endpoint to the initial velocity (the Jacobian required by Newton's method) is not approximated by finite differences, which become numerically singular for extreme shape parameters ($\alpha\gtrsim 10^4$). Instead, it is computed exactly by integrating the *variational equations* (the Jacobi field) alongside the geodesic itself, as an augmented ODE system. Each Newton step is damped by backtracking: if a full step worsens the endpoint mismatch, the step is repeatedly halved (up to $2^{-7}$) until it improves. Forward integration uses a fixed-step fourth-order Runge-Kutta scheme, with the number of steps (400-4000) scaled to a stiffness estimate based on the ratio $\max(\alpha)/\varphi(\min\alpha)$, since the metric becomes nearly degenerate at small shape values.

Stage 2: First-integral quadrature

If shooting fails to converge, the solver exploits two conserved quantities that hold along any geodesic of this metric (Eq. 3.5 of [19]). These first integrals reduce the boundary value problem to a system of two scalar equations in two unknowns: the total geodesic length $\rho$ and the minimum shape value $\alpha_{\min}$ attained along the path. Two geometrically distinct path types are considered: a direct branch, in which $\alpha$ varies monotonically between the endpoints, and a turnaround branch, in which $\alpha$ first decreases to $\alpha_{\min}$ and then increases. The integrals defining the equations contain a square-root singularity at the turning point $\alpha=\alpha_{\min}$. This is removed analytically by the substitution $u=\sqrt{\alpha-\alpha_{\min}}$ (equivalently, $\alpha=\alpha_{\min}+u^2$), so that $d\alpha=2u\,du$ cancels the $\frac{1}{\sqrt{\alpha-\alpha_{\min}}}$ factor and the integrand becomes regular. Here $u$ runs from 0 at the turning point $(\alpha=\alpha_{\min})$ to $\sqrt{\alpha_k-\alpha_{\min}}$ at an endpoint of shape $\alpha_k$. For the turnaround branch the integral is split at the turning point into two monotonic legs: one descending from the start shape down to $\alpha_{\min}$, one ascending from $\alpha_{\min}$ up to the end shape, each evaluated under this substitution. The resulting integrals are evaluated by adaptive Gauss-Kronrod quadrature (absolute tolerance $10^{-10}$, relative tolerance $10^{-8}$).

The two-equation system is solved with a trust-region nonlinear solver from multiple seed values of $\alpha_{\min}$ (spanning twelve orders of magnitude, plus a seed derived from the Poincaré approximation), with the unknowns transformed (log and logit) to enforce their natural bounds. A candidate solution is accepted only if its residual is below $5 \times 10^{-8}$; when both branches yield valid solutions, the one with the smaller residual (and, at parity, the shorter length) is retained. The geodesic path itself is then reconstructed pointwise by quadrature of the same integrals.

Stage 3: Collocation

For any remaining cases, the boundary value problem is solved directly with MATLAB R2025b's adaptive collocation solvers (bvp4c, falling back to bvp5c), initialized from a straight-line path between the endpoints.

2.4.3 Distance Evaluation and Output

Once the geodesic is found, the Fisher-Rao distance is its arc length. For solutions obtained by shooting or collocation, the path is integrated at high resolution and the local speed

$$\sqrt{\frac{\varphi(\alpha)}{\alpha}\left(\frac{d\alpha}{dt}\right)^2 + \alpha\left(\frac{d\eta}{dt}\right)^2} \tag{2}$$

is accumulated by trapezoidal quadrature; for the first-integral route, the length $\rho$ is itself one of the solved unknowns. The geodesic path is returned in the original (shape, scale) coordinates via the inverse transformation $\beta = e^{\eta}/\alpha$. Coincident input distributions (agreeing to a relative tolerance of $10^{-10}$) are detected up front and assigned zero distance.

### *2.5 Derived Metrics*

Cohort

Nineteen participants completed both conditions. One was excluded for poor-quality data collection that prevented reliable gamma fits, leaving 18 participants for all primary analyses. Analyses built on the windowed trajectory additionally required a minimum number of valid windows per condition. This retained 17 participants for the heavy-tail decomposition and 15 for the analysis using windowing.

Each participant performed one low-load and one high-load session. From each of those sessions, we derive two representations of the micromovement fluctuations. The static representation concatenates all speed segments and fits a single gamma distribution to the normalized micromovements, yielding one point ($\alpha$, $\beta$) per participant per condition. The windowed representation slides an 800-sample window (25-sample step) (approximately 3.3s with 0.1s step) across the concatenated raw speed series and fits a gamma PDF on the micromovements within each window. For a window to be considered valid, it must contain at least 15 micromovement peaks. For a subject to be considered valid, they must have at least 30 valid windows per condition. This yields an ordered sequence of shape values that traces the participant's path through the plane. The static and the windowed representations are not interchangeable, as we find unique results within each type of characterization.

For each participant we compute the Fisher-Rao distance between their low and high-cognitive load static points, $d_{FR}(\theta_{low}, \theta_{high})$. This is the summary of how far cognitive load moves a participant's fluctuation signature, on the parameterization-independent scale developed earlier.

The static shape parameter α conflates how often a participant's motor noise enters a heavy-tail regime with how deep that regime is. Therefore, we decompose the windowed shape sequence into two interpretable axes. For a threshold τ on α we define, per participant per condition, the frequency f(τ), which is a fraction of windows with < τ, and the severity v(τ), the mean of log(α) over those heavy-tailed windows. We sweep over all α and select τ which collects data below the 25th percentile of α, which occurs at approximately 11.8. Each participant x condition is then summarized as a point in (f, v) in a 2D fluctuation-regime plane. The interpretation of this is the following: how frequently does a participant venture into the heavy-tailed regime, and when they do enter it, how heavy does the tail get?

*2.6 Statistical Analysis*

All primary analyses use the n=18 baseline cohort, ignoring the one corrupted sample. The heavy-tail decomposition included participants with ≥ 30 valid windows per condition. Analysis requiring per-window convergence use n=15.

The per-participant Fisher-Rao low-vs-high distances are summarized across participants. To test whether load reduces between-subject variability in the static representation, we quantify the dispersion of the per-condition (α, β) cloud in two ways. On the Fisher-Rao metric itself, within each condition we compute the cohort centroid as the Fréchet mean (the point minimizing distances on a curved surface) of the per-subject (α, β) points under the FR metric. We define the between-subject radius as the root-mean-square FR distance from each subject to that centroid. In parameter coordinates, we compute the between-subject standard deviation of log α and of log β within each condition, working in log space because both parameters are positive and right-skewed. For each measure we summarize the load-induced change as the ratio of its low to high-load value, so that a ratio above one denotes contraction of the cloud under load.

We assess contraction inferentially in two ways. As the primary test, because the design is paired, we compare each subject's geodesic distance to its own-condition Fréchet mean between conditions with a Wilcoxon signed-rank test. Finally, we place a 95% confidence interval on each contraction ratio with a subject-level bootstrap. We hold the Fréchet means fixed at their full-sample estimates, which yields a deliberately narrow interval.

Two controls guard against the interpretation that the low-load cloud is merely noisier rather than genuinely more spread. First, we estimate a within-subject noise floor: for each subject and condition we split the session into two halves (first half vs. last half by segment, and odd vs. even segments), fit an independent gamma PDF to each half, and compute the FR distance between the two fits. Because halving the data inflates the sampling variance of each fit, this split-half distance is a conservative, upward-biased estimate of the within-subject noise. This split-half distance indexes the reliability of a single (α, β) estimate; we compare it across conditions with a paired test to confirm that within-subject reliability does not itself differ by load.

Second, we test for within-session nonstationarity by comparing the time-ordered split (first half vs. last half), which is sensitive to drift, warm-up, and fatigue, against the interleaved split (odd vs. even segments), which reflects sampling noise alone. A time-ordered distance reliably larger than the interleaved one would indicate drift, so that the dispersion attributed to between-subject differences is not an artifact of drift within a session. Finally, we place the between-subject radius on the same scale as this within-subject floor by taking their ratio per condition and report this radius-to-noise ratio explicitly, noting that under the proper FR metric the radius approaches this conservative floor in both conditions, so the static cloud sits near the resolution limit of a single-session estimate, and its endpoint cannot be cleanly separated from measurement noise. This is a limitation of the pooled static representation and a further reason the windowed analysis carries the convergence claim.

Because the static representation pools every micromovement into a single gamma fit, it can overstate between-subject convergence: a subject who only intermittently enters the heavy-tail regime is summarized by one shape value that absorbs both regimes. To check this, we compare each subject's pooled static shape against the geometric mean of their windowed shape sequence, using a paired t-test per condition. A systematic static-minus-windowed difference, and one that differs in size between conditions, would indicate that pooling the spikes biases the static shape and that the apparent static collapse would partly be an aggregation artifact. We therefore also recompute the between-subject SD(log $\alpha$) contraction ratio from the windowed summaries and report it alongside the static ratio. The windowed contraction, exchanging the pooling bias for a potentially smaller bias from a small sample, motivates the windowed decomposition that follows, where the statistical evidence for convergence rests on the baseline-predicts-change analysis rather than on a variance ratio.

We test the load effect on each axis of the (f, v) decomposition separately. For frequency and for severity we compare the low and high-load values across participants with a paired t-test (one-sided for frequency, under the directional hypothesis that load reduces how often the heavy-tail regime is entered; two-sided for severity). To characterize how each effect depends on the heavy-tail threshold rather than committing to the single $\tau = 11.8$ cut, we repeat both tests across the full sweep $\tau \in [3, 18]$ and report the range over which they hold. Convergence of the cloud is indexed by the between-participant standard-deviation ratio $SD_{low} / SD_{high}$ on each axis, again reported across the sweep. Finally, to test whether participants move toward a common point rather than shifting uniformly, we correlate each participant's baseline (low-load) value with their load-induced change ($\Delta$ = high − low) using Pearson's r, for frequency and for severity; a strong negative correlation indicates that participants who start farthest from the attractor move most toward it.

The baseline-predicts-change correlation is mechanically biased toward negative values: the baseline frequency (or severity) appears on the predictor axis and again, with a negative sign, inside the change ($\Delta$ = high − low), and because each per-participant baseline is estimated from a finite number of windows it carries sampling noise that regresses toward the mean on the second measurement. A negative correlation would therefore arise even if load had no effect. To quantify this artifact and test whether the observed correlation exceeds it, we construct an exchangeability null. For each participant we pool their low- and high-load windowed shape values and randomly re-partition them into a pseudo-low and a pseudo-high set of the same sizes as that participant's true conditions, so the two pseudo-conditions are statistical twins differing only by the random shuffling. We then recompute the baseline-predicts-change correlation across participants. Repeating this many times yields the distribution of correlations produced by the mechanical coupling and the real measurement noise at the true sample sizes, but with any genuine load effect removed. We compare the observed correlation against this null distribution, reporting the null mean as the artifact floor and the proportion of null draws as extreme as the observed value (the permutation p-value).

As a second, coupling-free check we report Oldham's correlation [21] between the change and the average of the two conditions, (low + high)/2, rather than the baseline. This quantity is free of the baseline-on-both-axes coupling, and its covariance with the change reduces to half the difference of the two conditions' variances; it is therefore negative only when the high-load cloud is genuinely less dispersed than the low-load cloud, providing independent confirmation of contraction.

## 3. Results

Cognitive load moved every participant's micromovement distribution: the Fisher-Rao distance between low- and high-load static (α, β) fits had a median of 0.59 (range 0.05-1.22; Fig. 2). But the displacement had no shared direction. α increased under load in only 9 of 18 participants (sign test $p = 1.000$), and paired t-tests on log α and log β were both null ($p = 0.85$, 0.88). The cohort Fréchet means were near-coincident, (11.09, 0.055) low vs. (10.54, 0.058) high. The cloud tended to contract. The Fisher-Rao Fréchet radius fell 1.30× (0.524 → 0.402), with the parameter-coordinate spread agreeing (SD log α 1.55×), but the trend did not reach significance (signed-rank one-sided $p = 0.075$; bootstrap 95% CI on the ratio [0.94, 1.96], containing one). Split-half reliability was identical across conditions ($p = 0.998$), so the low-load cloud is not merely noisier; but the radius sits near the within-participant noise bound, so the pooled static estimate is at its resolving limit. This motivates the windowed decomposition.

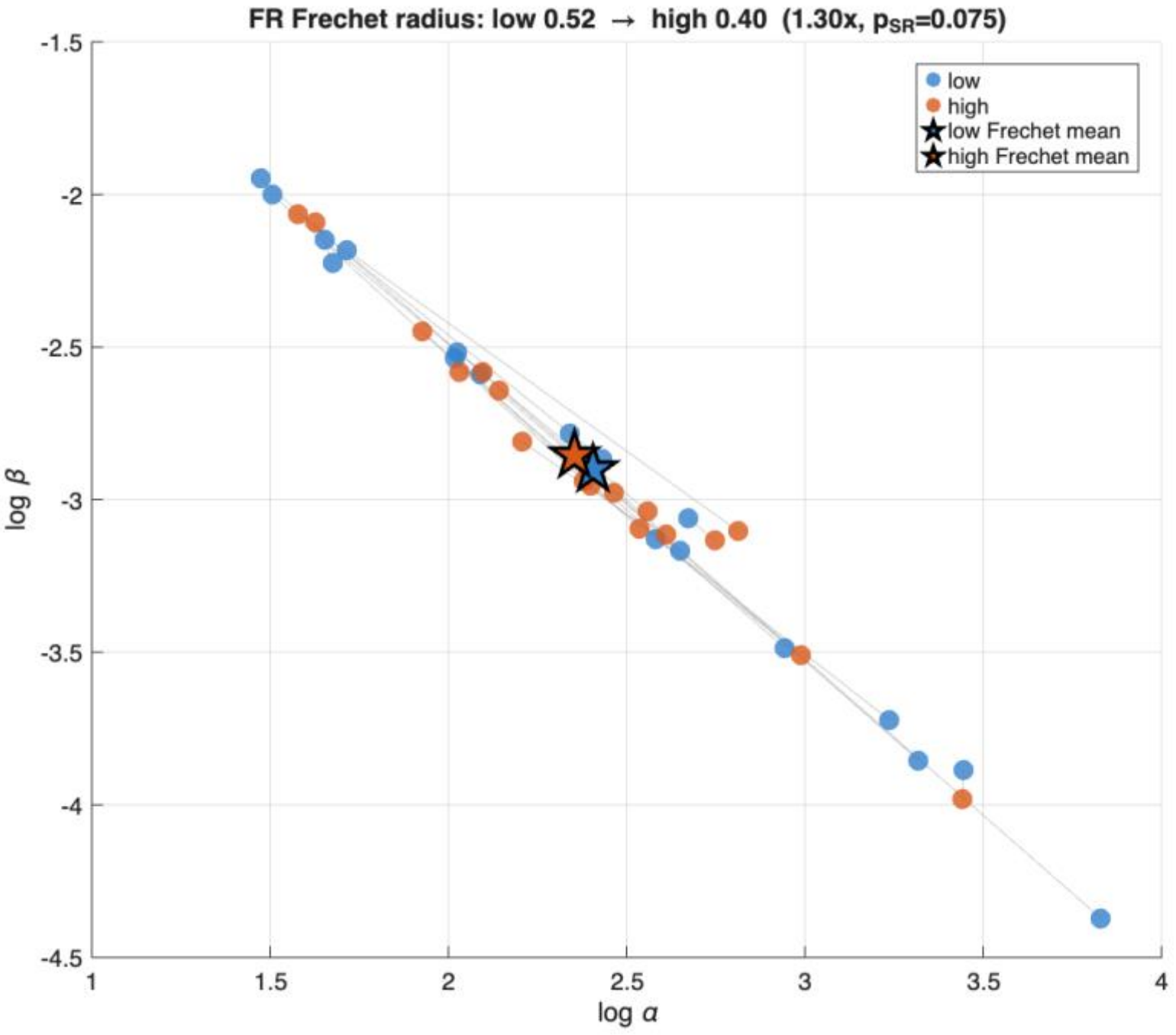

Figure 2 Pooled analysis of Fréchet radius does not show a significant contraction between low and high cognitive load.

Pooling every micromovement into one fit overstated convergence: the pooled static shape sat below the geometric mean of the windowed sequence (paired t, $p = 0.01$ low, $p < 0.001$ high), and the between-participant SD(log α) contraction was correspondingly weaker when measured from the windowed summaries (1.23×) than from the static fits (1.55×). We therefore analyzed the windowed shape sequence directly, splitting it at $\tau = 11.8$ (the pooled-α 25th percentile) into how often a participant entered the heavy-tail regime (frequency) and how deep it ran when they did (severity). This pipeline is demonstrated in Figure 3, below.

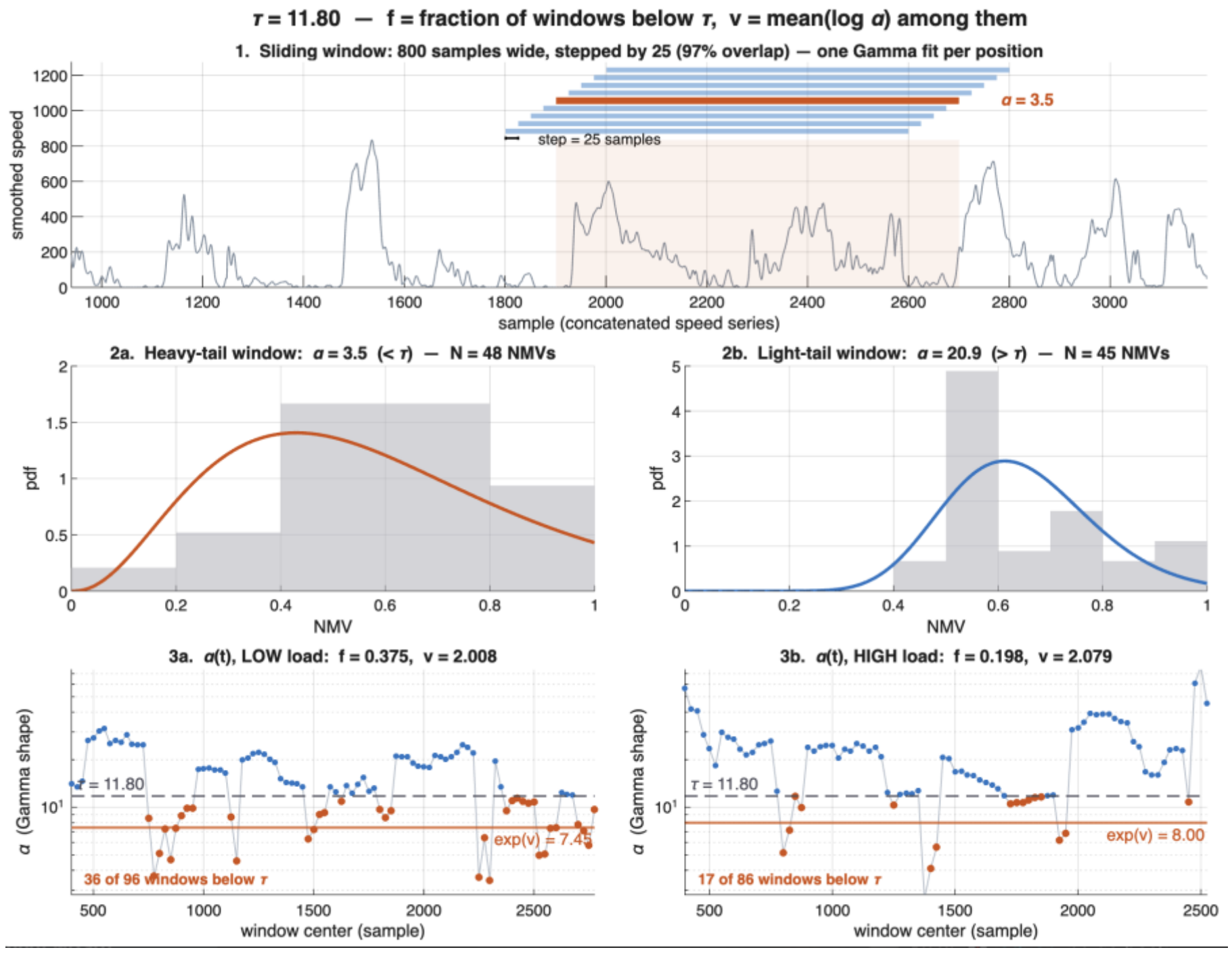

Figure 3 - Pipeline for α frequency/severity decomposition.

The dominant effect was convergence toward a common point rather than a uniform shift. A participant's baseline level strongly predicted their load-induced change on both axes: $r = -0.89$ for frequency and -0.81 for severity (threshold-robust across the sweep, frequency $r \in [-0.79, -0.91]$), so participants who began farthest from the attractor moved most toward it, with the high-load cloud settling near $(f^*, v^*) \approx (0.22, 1.78)$ (Fig. 4). Because correlating a baseline against its own change is mechanically biased negative, we tested these against an exchangeability null: the artifact alone produced only $r \approx -0.31$ (frequency) and -0.40 (severity), far short of the observed values (permutation $p = 0.0002$ and 0.0024, Fig. 5), and Oldham's coupling-free correlation remained negative (-0.51, -0.43). The convergence is therefore real, not a regression-to-the-mean artifact.

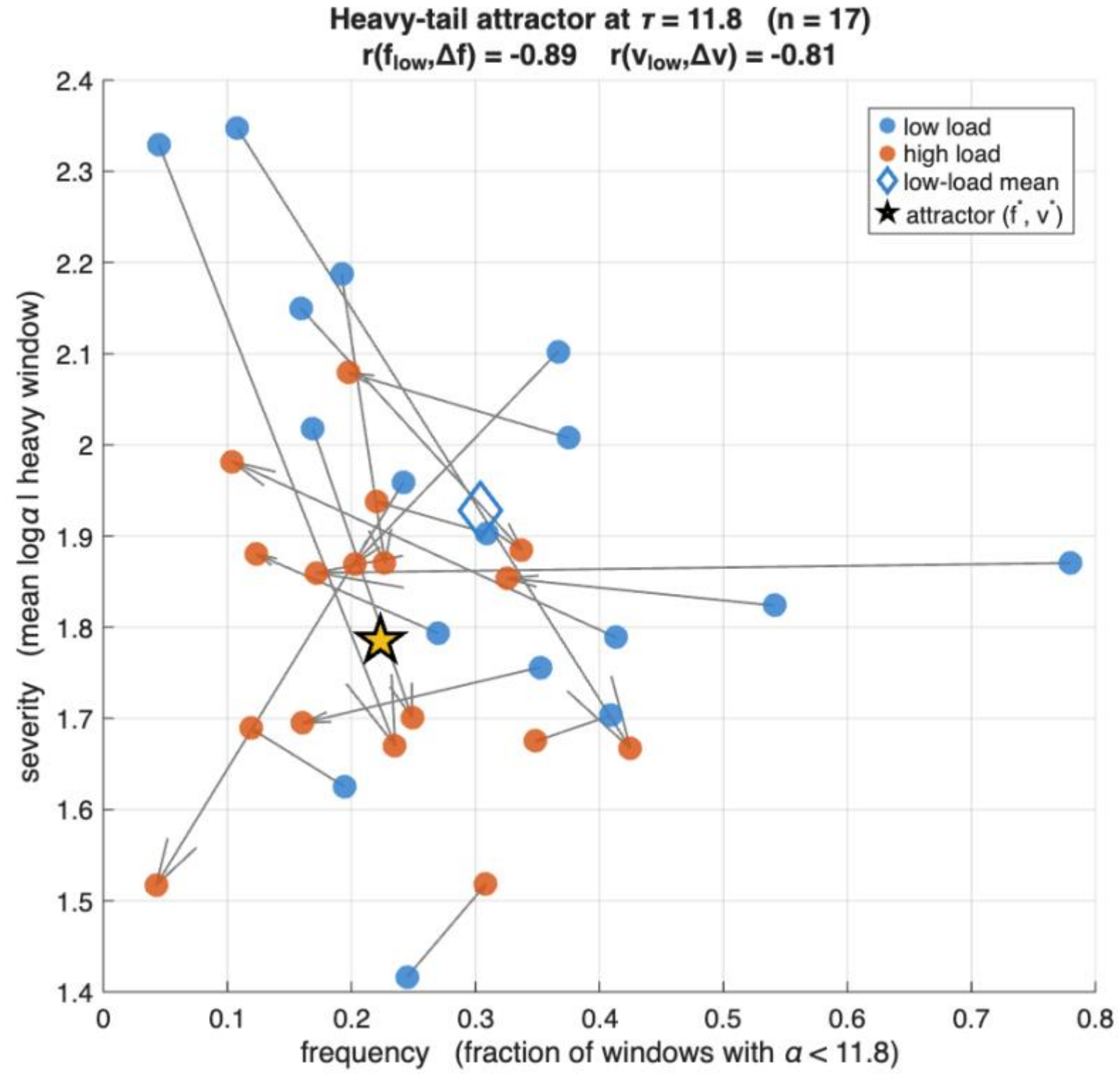

Figure 4 - High Load Makes Frequency and Severity of Windows with Heavy Tails More Homogenous. Specifically, heavy tail frequency and severity between low and high cognitive load is negatively correlated

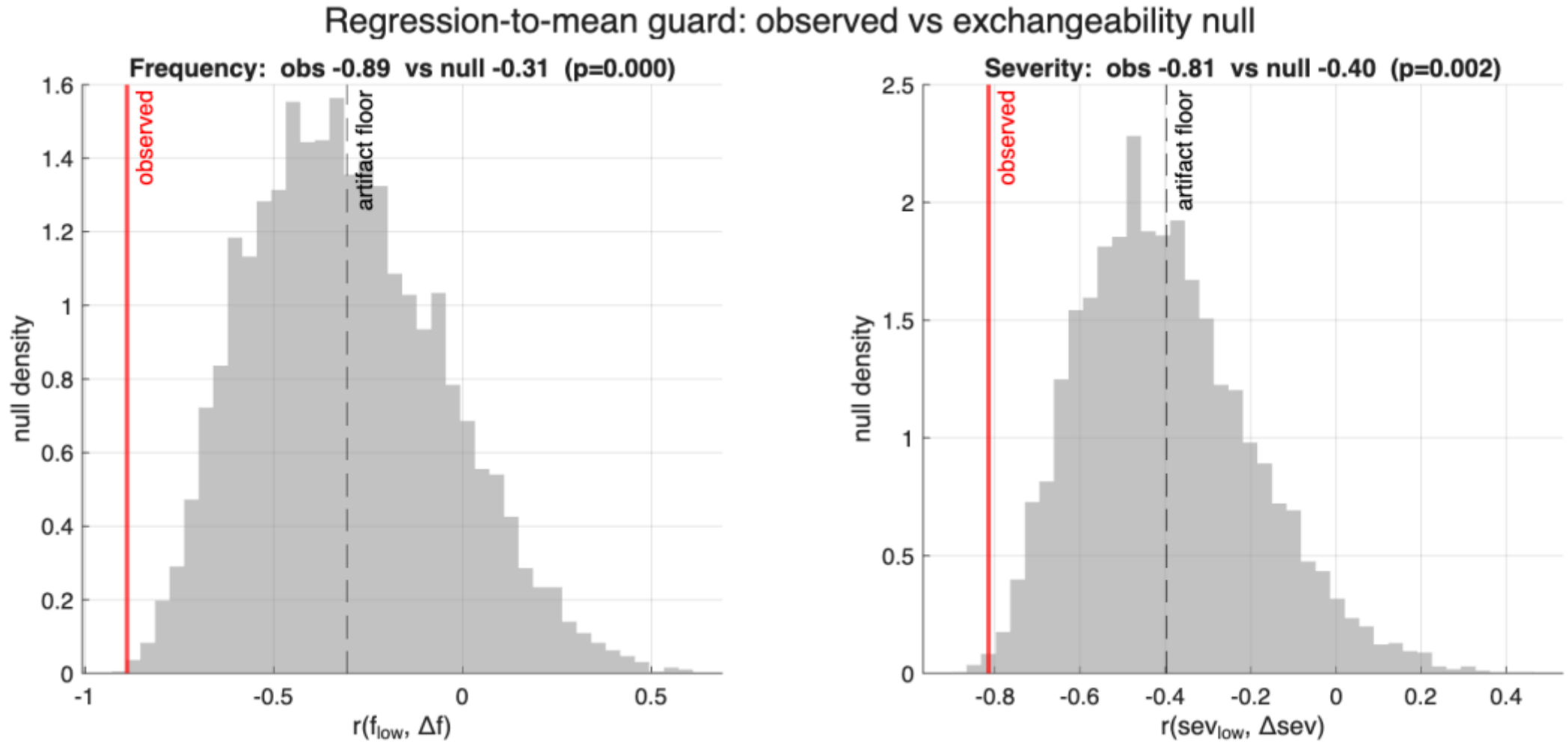

Figure 5 - Change in frequency and severity of heavy Tails is probably not regression to the mean. When you shuffle the high cognitive load values within similar windows to simulate a null effect, the observed change in correlation is significantly different than that distribution

The mean shifts were consistent with this picture but secondary. The heavy-tail regime deepened when entered. Among heavy windows, mean log α fell from 1.92 (low) to 1.77 (high), $p = 0.03$, significant across $\tau \in [10, 11.5]$, while the frequency of entering it dropped from 0.30 to 0.22 but only marginally at this threshold (one-sided $p \approx 0.07$, strengthening to $p = 0.03$ for $\alpha \geq 15$). The effect of load is thus better described as a contraction of between-participant variability onto a shared operating point, accompanied by a significant deepening of the tail, than as a uniform population-wide shift.

## 4. Discussion

This paper's main contributions are a measurement instrument and a simple result that the analysis enables. First, we adopt the Fisher-Rao Distance as a coordinate-free ruler on the gamma plane, such that distances between estimates are both robust to parameterization (shape and scale vs shape and rate), as well as including the inherent anisotropy of the space. Second, we supply the machinery to compute these distances, which will be made available in both MATLAB R2025b and Python 3.11, building on the geodesic formulation of Reverter and Oller [19]. This solver implements the three-stage solution discussed in the previous section and has been verified against Reverter and Oller's reference tables, in addition to additional theoretic invariants documented in the repository. The solver converges across the full range of shape parameters produced by the data, including the large-α regime where a naïve shooting method fails. This allowed us to make metric-aware calculations on the cohort's data on the actual manifold rather than in flat space. This makes it possible to ask whether a population of distributions has moved or tightened: a question that has no canonical answer without a canonical sense of taking a measurement.

The TMT application is a concrete demonstration of the ruler, rather than the principal claim, and it is informative because it is precisely the regime where summary statistics and geometry-agnostic methods fail. The means of the distributions were not the measure of interest, but in fact the spread of the distributions. Cognitive load displaced every participant's fluctuation signature: the Fisher-Rao distance between low- and high-load static fits had a median of 0.59 and ranged from 0.05 to 1.22, yet the displacement had no shared direction. The shape parameter increased under load in only 9 of 18 participants (sign test $p = 1.000$), paired t-test on $\log \alpha$ and $\log \beta$ were both null ($p = 0.85$ and 0.88), and cohort Fréchet means were nearly identical at low load versus high load.

The novelty of this work is primarily a combination of the micromovement construct on the gamma plane with the numerical calculation of the geodesic on the gamma manifold. Prior applications of the gamma-plane representation used standard hypothesis testing such as the Wilcoxon ranksum test to detect differences, but non-parametric tests like this ignore the intrinsic curvature of the space the parameters lie on, admittedly on purpose, hence non-parametric. An unstated consequence of this, though, is that these types of tests can only detect the pairwise differences in each person, while being agnostic to the population-level compression of the dataset variance.

One methodological result that is particularly useful here is the pooled fitting null result. On inspection, when comparing the windowed gamma fits to the pooled fits, we notice that the pooled estimate represents approximately none of the data accurately, as the shapes span into two shape regimes: symmetric and heavy-tailed distributions, and splitting the difference loses the distinct signature each person has. A second limitation on the static representation is resolution. The Fisher-Rao Fréchet radius sits at or below a deliberately conservative within-participant noise bound in both conditions (radius-to-noise 0.82 low, 0.59 high). Split-half reliability was statistically identical across conditions ($p= 0.998$) and we found no within-session drift ($p=0.27$ and 0.84), so the low-load cloud is not simply noisier than the high-load one, but between-participant structure in the pooled static estimate is close enough to the noise it can resolve that the static contraction is not sufficient evidence on its own.

When considering the windowed analysis, the effect of cognitive load is better described as a contraction rather than a translation. Labeling the shapes as different classes of heavy-tailed-ness allowed us to compute the relationship between the low-load shape and the change of shift frequency into the heavy-tailed regime, as well as low-load shape versus the intensity of the shape change, which each had strong negative correlations ($r = -0.89$ and $-0.81$, respectively). This significance finding was not dependent on the threshold of α. As stated in the results, this finding is unlikely to be regression to the mean, as demonstrated with the permutation null and Oldham's method.

The reading we favor is that the fluctuation signature is something spare capacity affords. When executive resources are committed to maintaining and switching between two ordered sets, motor output falls back towards a common operating regime, and individuality in the fluctuation signature is among the first things given up. This is a different claim from the familiar one that load adds noise. That interpretation is more amenable to novel and more kinematically difficult tasks where execution itself is a limiting factor, unlike a simple drawing task. This reading that structured variability is functional is consistent with [7]: if the structure in a participant's fluctuation reflects an actively maintained control policy, then the resource cost of maintaining it is what a secondary cognitive demand competes for. This is a proof of concept in 18 healthy adults under an experimental load manipulation. We make no clinical claims here, but the Fisher-Rao tool remains a viable option for quantifying shifts via cognitive load in clinical cases. Comparing that against an established baseline is critical for quantifying efficacy of interventions.

Limitations of this study include sample size and condition order. The cohort is small, with only 18 total participants. Only a subset of 17 were used for windowed heavy-tail decomposition. The ages, motor capability, and cognitive capability are not a representative sample. Additionally, the condition order was not randomized, so the contraction finding here could be limited to this specific ordering of the task. One major consideration hedges against this shortcoming, however. We found no within-session drift under a time-ordered split ($p=0.27$ and 0.84). This is not a substitute for counterbalancing, however, and future work should resolve this confound for both the current analysis for replication as well as for any future analysis using dTMT-A and dTMT-B.

Nothing in the ruler is specific to stylus kinematics. The same construction applies to any positive-valued fluctuation series adequately described by a gamma family: inter-stride intervals, tremor amplitude, speech envelope peaks, and the invariance that motivates it becomes more valuable, not less, as cohorts become more heterogeneous in body size and anatomy, since it removes the dependence on units and coordinates that makes raw kinematic scalars difficult to compare across people. Two questions follow directly from the present result and are testable with the same apparatus. First, whether a participant's baseline distance from the common operating point indexes something like reserve and predicts how much load it takes to move them. Second, whether populations with reduced executive capacity already sit near that operating point at baseline. If that is true, it would make the absence of individuality under low load the measurement of interest and would require exactly the kind of parameterization-independent, within- and between-person comparable distance developed here.

**Author Contributions:** Conceptualization, J.V. and E.B.T.; methodology, J.V.; software, J.V.; validation, J.V.; formal analysis, J.V.; investigation, J.V.; data curation, J.V.; writing—original draft preparation, J.V.; writing—review and editing, J.V. and E.B.T.; visualization, J.V.; resources, E.B.T.; funding acquisition, E.B.T. All authors have read and agreed to the published version of the manuscript.

**Funding:** This research was funded by The Nancy Lurie Mark Family Foundation Career Continuation Award to EBT and by the Rutgers TA-ship award to JV.

**Institutional Review Board Statement:** Recruitment and consent procedures were approved by IRB protocol # Pro2020000154 .

**Informed Consent Statement:** Informed consent was obtained from all subjects involved in the study

**Data Availability Statement:** Due to the identifiability of the cohort, we will not be making the data available. However, we will make the data acquisition app available at a later time.

**Acknowledgments:** During the preparation of this manuscript/study, the author(s) used Claude Opus 5 for the purposes of line-level writing edits and initial generation of plotting scripts. The authors have reviewed and edited the output and take full responsibility for the content of this publication.

**Conflicts of Interest**: Rutgers University owns a patent on some of the technology used in the paradigm and products being studied. Some of the analytical approaches are protected under granted patents by the US Patents and Trademark Office including US10176299B2– Methods for the diagnosis and treatment of neurological disorders; US20170344706A1 – Systems and methods for tracking neurodevelopment disorders; US20170340261A1 – System and method for measuring physiologically relevant motion; and US10786192B2 – System and method for determining amount of volition in a subject. JV and EBT have disclosed the present technology to Rutgers University Office of Innovation and Venture in consideration for a provisional patent.

## Abbreviations

The following abbreviations are used in this manuscript:

| | |
|---|---|
| TMT | Trail Making Task |
| dTMT | Digitized Trail Making Task |
| FR | Fisher-Rao |
| MoCA | Montreal Cognitive Assessment |
| MM | Micromovements |
| MLE | Maximum Likelihood Estimation |
| MCI | Mild Cognitive Impairment |
| ODE | Ordinary Differential Equation |
| BVP | Boundary Value Problem |

# Appendix A

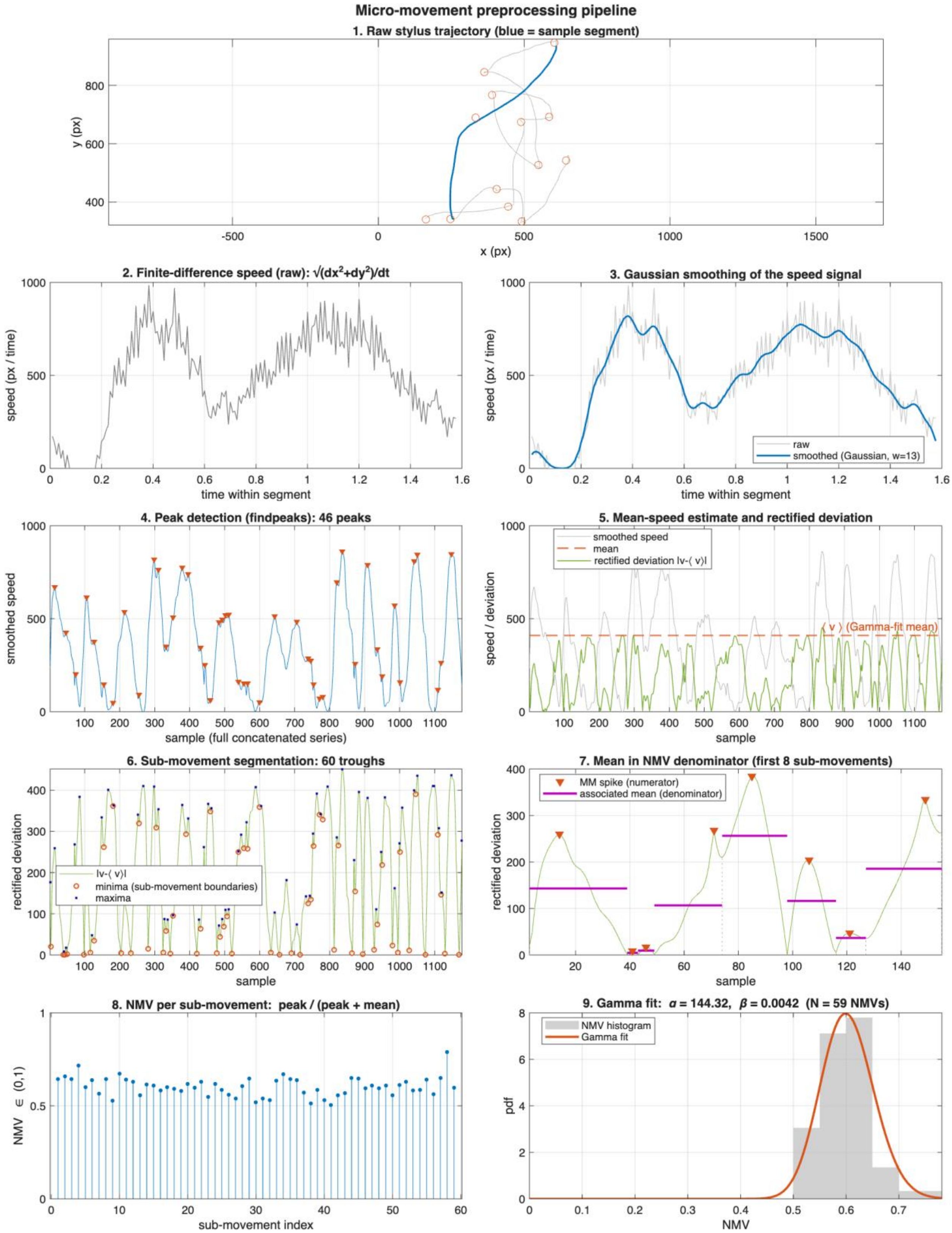

Figure A1 - Each panel of the Micromovement Pipeline visualizes the sequence of the calculation